\documentclass{aa}  

\usepackage{graphicx}
\usepackage{txfonts}
\usepackage[normalem]{ulem}

\usepackage{color}           

\usepackage{natbib}
\bibpunct{(}{)}{;}{a}{}{,} 

\begin{document}

\title{Electron dynamics in small magnetospheres}
\subtitle{insights from global fully-kinetic plasma simulations of planet Mercury}

\author{ Federico Lavorenti
          \inst{1,2}\thanks{federico.lavorenti@oca.eu}
          ,
         Pierre Henri
          \inst{1,3}
          ,
         Francesco Califano
          \inst{2}
          ,
         Jan Deca
          \inst{4,5,6}
          ,
         Sae Aizawa
          \inst{7}
          ,
         Nicolas Andr\'e
          \inst{7}
          and
         Johannes Benkhoff
          \inst{8}
          }
          
 \institute{ 
 Laboratoire Lagrange, Observatoire de la C\^{o}te d’Azur, Universit\'e C\^{o}te d’Azur, CNRS, Nice, France
 \and
 Dipartimento di Fisica "E. Fermi", Università di Pisa, Pisa, Italy
 \and
 LPC2E, CNRS, Univ. d'Orl\'eans, OSUC, CNES, Orl\'eans, France
 \and
 LASP, University of Colorado Boulder, Boulder, CO, USA
 \and
 Institute for Modeling Plasma, Atmospheres and Cosmic Dust, NASA/SSERVI, Silicon Valley, CA, USA
 \and 
 LATMOS, Université de Versailles à Saint Quentin, Guyancourt, France 
 \and
 IRAP, CNRS-CNES-UPS, Toulouse, France
 \and
 ESA/ESTEC, Keplerlaan 1, 2200 AG Noordwijk, The Netherlands
              }

\titlerunning{Global fully-kinetic plasma simulations of Mercury}

\authorrunning{Lavorenti F., et al.}

 \date{Received ?, 2022; accepted ?}

\abstract
{The planet Mercury possesses a small but highly dynamic magnetosphere in which the role and dynamics of electrons are still largely unknown.}
{We aim at modeling the global dynamics of solar wind electrons impinging on Mercury's magnetosphere. Particular relevance is given to local acceleration processes and the global circulation patterns.}
{The goals of this work are pursued by means of three-dimensional, fully kinetic particle-in-cell simulations modeling the interaction of the solar wind with the Hermean magnetosphere. This method allows a self-consistent representation of the plasma dynamics from the large planetary scale down to the electron kinetic scale. Numerical simulations are carried out using two different solar wind conditions: purely northward or purely southward interplanetary magnetic field direction.}
{We find a high plasma current (of the order of few $\mu$A/m$^2$) flowing at the magnetospheric boundaries (bow shock and magnetopause) dominated by electrons. This current is driven by the small-scale electron physics resolved in our model.
Furthermore, we observe strong electron acceleration up to tens of keV as a consequence of magnetic reconnection when the interplanetary magnetic field is directed southward. Such energetic electrons are partially trapped in the dipolar magnetic field of the planet mainly at nightside. 
Finally, by studying the distribution of electrons in our simulations along Mariner10 and BepiColombo first-Mercury-flyby trajectories, we propose that both spacecraft observed this energetic quasi-trapped electron population around closest approach.}
{}


\maketitle

\section{Introduction}\label{sec:Introduction}

Mercury is one of the least explored planet of the solar system. In decades of space exploration, only two missions have been devoted to the innermost planet of the solar system. 

The NASA Mariner10 mission in the 1970's provided a snapshot of the Hermean environment with its three flybys~\citep{Russell1988}. These observations showed the presence of a planetary magnetic field~\citep{Ness1974} and of a structured plasma environment~\citep{Ogilvie1977}. In particular, Mariner10 was able to perform measurements of the core of the electron distribution function (in the range 13.4-687 eV) during most of its orbit inside the Hermean magnetosphere~\citep{Christon1987}. However, Mariner10 was not able to observe ions due to a technical failure~\citep{Ogilvie1974}.

The NASA MESSENGER mission deeply extended our knowledge of the Hermean environment during its four years of orbital observations~\citep{Solomon2018}. This wealth of \textit{in situ} observations enabled the first systematic studies of the interaction between the solar wind plasma and Mercury's magnetized environment~\citep[and references therein]{Raines2015}. These studies put in evidence the highly dynamical character of Mercury's plasma environment and shed light on the tight coupling between the solar wind, magnetosphere, exosphere and surface of Mercury. The mission addressed several plasma processes occurring at the global planetary scale (of the order of 2400 km) and down to the ion kinetic scale (of the order of 100 km)~\citep{Boardsen2012,Raines2014,Gershman2014,Gershman2015,Schmid2021}. However, given the instrumental constraints of the mission, MESSENGER did not address the plasma processes ongoing at the electron scale. The instrumental suite onboard MESSENGER provided observations of electrons above $\sim$ 10 keV, thus missing the core of the distribution function, and it could not provide plasma waves observations below the typical ion frequency range (of the order of few Hz). 

The ESA/JAXA BepiColombo mission has been designed to continue building upon the legacy left by MESSENGER~\citep{Benkhoff2021}.
BepiColombo is now cruising to Mercury (arrival expected in 2025) and performed its first Mercury flyby the 1st of October 2021. The novelty of this mission resides in (i) its two-spacecraft nature (BepiColombo is composed of one Mercury Planetary Orbiter, MPO, and of one magnetospheric orbiter, nicknamed Mio) and in (ii) its advanced instrumental suite for environmental studies enabling observations down to the electron scale~\citep{Milillo2020}. BepiColombo is the first mission able to provide a simultaneous multi-point picture of the Hermean environment from the global planetary scale down to the electron scale (of the order of 1 km). In order to both analyze and optimally plan such novel observations, global numerical models of the Hermean environment with resolution down to the electron scale are needed.

To date, global numerical models of Mercury's plasma environment have been mostly limited to magnetohydrodynamic (MHD)~\citep{Kabin2000,Ip2002,Yagi2010,Pantellini2015,Jia2015,Jia2019}, multi-fluid~\citep{Kidder2008,Benna2010,Dong2019} and hybrid (meaning kinetic ions and fluid mass-less electrons)~\citep{Kallio2003,Travnicek2007,Travnicek2009,Travnicek2010,Richer2012,Muller2012,Hercik2013,Hercik2016,Fatemi2018,Fatemi2020,Exner2018,Exner2020} models. These models reproduce appropriately the plasma dynamics down to the ion kinetic scale and have been extensively and successfully used in support to the Mariner10 and MESSENGER missions. However, to interpret the forthcoming new observations of the BepiColombo mission, more refined models able to self-consistently include electron kinetic physics are required.

Recently, a first attempt to locally embed electron kinetic physics in a global MHD simulation has been carried out by~\citet{Chen2019}. Their model can be used to study the role of electrons in a precise subset of the global magnetosphere (for instance, the magnetotail), on the other hand it cannot reproduce dynamical processes encompassing the global magnetosphere system such as, for example, the global electron circulation around the planet~\citep{Walsh2013}. In order to overcome this limitation and to include self-consistently both ion and electron physics, in this work, we study the interaction between the solar wind and the Hermean plasma environment using a global fully-kinetic model. 
A similar approach has been presented by~\citet{Lapenta2022} using the results of a global hybrid simulation to initialize a fully-kinetic one. In that work, the authors found that the feedback of kinetic electrons does not affect the large scale structure of the magnetosphere. Nonetheless, on the small scale, they found (i) thinner magnetospheric boundaries with higher currents and (ii) more efficient energization processes in the magnetosphere. The results of~\citet{Lapenta2022} point out the importance of a kinetic model for electrons in global simulations of planetary magnetospheres, a goal today attainable thanks to the increasing computational power of nowadays HPC facilities.

Although only a few  \textit{in situ} electron observations are available at Mercury, some hints on their global dynamics have been discussed in past works using Mariner10~\citep{Christon1987} and MESSENGER  data~\citep{Ho2012,Ho2016,Baker2016,Dewey2018}. One of the most significant outcomes of these observations is the presence of a quasi-stable high-energy (up to tens of keV) electron population inside the Hermean magnetosphere, mainly observed in the night post-midnight sector (local time 0-6~h). 
According to those observations, the Hermean environment is significantly populated by electrons with energies up to tens of keV. In this work, we investigate the origin of these energetic electrons and their loss mechanism inside the Hermean magnetosphere by means of global fully-kinetic simulations.


The paper is organized as follows: 
in Sec.~\ref{sec:Model} we present our fully-kinetic model, in Sec.~\ref{sec:Results_large} we analyze the large scale structure obtained from our simulations and validate our model against nominal bow shock and magnetopause shapes and positions at Mercury,
in Sec.~\ref{sec:Result_electrons} we focus on our simulations results regarding electron energization and circulation in the Hermean magnetosphere,
in Sec.~\ref{sec:Result_comparison-data} we build and discuss synthetic electron energy spectra obtained from our simulations along Mariner10 and BepiColombo trajectories during their first Mercury flybys,
in Sec.~\ref{sec:Sec4} we discuss the results and conclude the paper.

\section{The model: fully-kinetic global simulations}\label{sec:Model}
The simulations presented in this work are done using the semi-implicit, fully kinetic particle-in-cell (PIC) code \texttt{iPIC3D} \citep{Markidis2010}. It solves the Vlasov-Maxwell system of equations for both ions and electrons by discretizing the distribution function of both species using macro-particles. We present two different simulations, namely RunN and RunS, characterized by a purely northward or southward interplanetary magnetic field (IMF), respectively. In the following, we use the Mercury-centered Solar Orbital (MSO) reference frame, where the x-axis points from the planet center to the sun, the z-axis is anti-parallel to Mercury's magnetic dipole and the y-axis points from dawn to dusk. 

The simulations use a three-dimensional cartesian geometry, a simulation domain $(L_x,L_y,L_z)$ = (-9:+6,$\pm$6,$\pm$6) $R$ (here R is the planet radius) in MSO coordinates, divided into $(N_x,N_y,N_z)$ = (960,768,768) grid cells. The simulations are initialized with $64$ macro-particles per cell per species. The time step is $dt$ = 1.4 ms, much smaller than the electron gyro-period ($\tau_{ce} \equiv 2\pi/\omega_{ce} \equiv 2\pi m_e c/e B_{_{SW}}$ = 31.5 ms). The grid spacing is $dx$ = $dy$ = $dz$ = 0.015 $R$ = 1.5 $\rho_e$, where $\rho_e \equiv c\sqrt{T_{e,_{SW}} m_e}/e B_{_{SW}}$ is the electron gyroradius. One important feature of implicit PIC codes is that, even if the grid spacing is of the order of the electron gyroradius $\rho_e \lesssim dx < \rho_i$, the code correctly captures the sub-grid electron kinetic physics as long as the condition 0.1 $< v_{the}dt/dx <$ 1 is fulfilled~\citep{Brackbill1982}. In our case $v_{the}dt/dx$ = 0.18.
However, since sub-grid oscillations on the electron scale tend to accumulate energy at the grid scale, a smoothing technique is required to avoid numerical instabilities at the grid scale (see~\citet{Toth2017} for details on the filter used; in our case the filter is applied two times per cycle with a factor $\alpha=0.5$).

The simulation setup includes (i) the solar wind plasma (uniform density, magnetic field and temperature), (ii) the spherical planet centered at $x,y,z=(0,0,0)$ with radius $R$, and (iii) the dipolar magnetic field centered in $x,y,z=(0,0,0.2)$ $R$ with intensity 200 nT$/R^3$.

Given that we simulate an infinite system using a finite numerical box, particular care must be taken when imposing the external boundary conditions. Because the planet interaction region is sufficiently far (of the order of 5 planet radii) from the external boundaries of the box, we populate the last cells of the boundaries with solar wind plasma (an exception is made for the antisunward boundary from which plasma is only allowed to exit the box). To smooth out the electromagnetic fields fluctuations close to the boundaries, we employ a LSAL (linear simple absorption layer) method~\citep{Berendeev2018} on the 10 outermost cells of the simulation domain. As discussed in~\citet{Berendeev2018}, this method avoids wave reflection and ensures numerical stability at the boundaries. With this method, the magnetic and electric field values in the outermost boundary cells are equal to their corresponding solar wind values, $\bf{B_{_{SW}}}$ and $\bf{E_{_{SW}}}$. Since in the solar wind the plasma frozen-in condition holds, the electric field is given by $\bf{E_{_{SW}}}=-\bf{V_{_{SW}}}\times \bf{B_{_{SW}}}$. A divergence cleaning routine for the magnetic field ensures that $\text{div}(\bf{B})=0$ in the absorbing layer.

Particular care must also be taken in handling the plasma interacting directly with the planet. This is especially true for models, such as fully-kinetic ones, that do not impose quasi-neutrality. In order to avoid the generation of spurious net charge density on the planet surface, the macro-particles falling into the planet are removed from the simulation using a charge-balanced method.
This method ensures that (i) the same amount of positive and negative charged macro-particles is removed from the simulation box at the planet surface boundary, (ii) the excess of electron flux at the planetary surface is expelled back radially.
This inner boundary condition mimics the {sub-grid} interaction between precipitating plasma and the planet surface ongoing on length scales {of few Debye lengths}.

In both simulations, the solar wind plasma is initialized with density $n_{_{SW}}$ = 30 cm$^{-3}$, velocity $\bf{V_{_{SW}}}$ = (-400,0,0) km/s, magnetic field $\bf{B_{_{SW}}}$ = (0,0,$\pm$20) nT and temperature $T_{i,_{SW}}$ = $T_{e,_{SW}}$ = 21.5 eV. The ram pressure is $P_{ram} \equiv n_{_{SW}}V^2_{_{SW}}$ = 8.2 nPa, and the plasma beta is $\beta \equiv 8\pi n_{_{SW}}T_{_{SW}}/B^2_{_{SW}}$ = 1.3.
The only parameter that is different between the two runs (RunN and RunS) is the direction of the magnetic field $\bf{B_{_{SW}}}$. These parameters are representative of those expected in the  solar wind at Mercury at aphelion~\citep{James2017,Sarantos2007} and similar to those used in~\citet{Aizawa2021}.

Once the interaction between the solar wind flow and the planetary magnetic field reaches a quasi steady-state, we stop the simulation. Such time scale corresponds approximately to the solar wind ballistic time to cross the simulation box, $T\approx L_x/V_{x,_{SW}}\approx 10$ s (corresponding to approximately 7000 numerical cycles). 

In order to keep computational resources manageable, we choose to reduce (i) the ion-to-electron mass ratio $m_i/m_e$ = 100 (from its real value of 1836), (ii) the electron plasma-to-cyclotron frequency ratio $\omega_{pe}/\omega_{ce}$ = 17.8 (from its real value of 83) and (iii) the planet radius $R$ = 230 km (from its real value of 2440 km).
With these values, we get a hierarchy of scale lengths $R$ = 10 $\rho_i$ = 100 $\rho_e$ that -- although compressed -- keeps a sufficiently large separation between planetary, ion and electron kinetic scales.
The first and second rescalings represent a standard procedure in plasma fully-kinetic simulations, see e.g.~\citet{Bret2010}. The third rescaling is analogous to what proposed and validated in~\citet{Toth2017} for the case of the Earth, and also discussed extensively in~\citet{Markidis2021}.
Moreover, this planet rescaling approach was already adopted in past works using global hybrid models in support to MESSENGER observations~\citep{Travnicek2007,Travnicek2009,Travnicek2010} and recently in a global fully-kinetic model~\citep{Lapenta2022}.

In the next section, we validate our scaled-down model against the nominal shapes and positions of the magnetosphere boundaries. A broader discussion on the impact of these rescalings on our results is postponed to Sec.~\ref{sec:Sec4}.

\section{Overview of the large scale structure: magnetosphere boundaries identification}\label{sec:Results_large}

\begin{figure*}
    \centering
    \includegraphics[width=\linewidth]{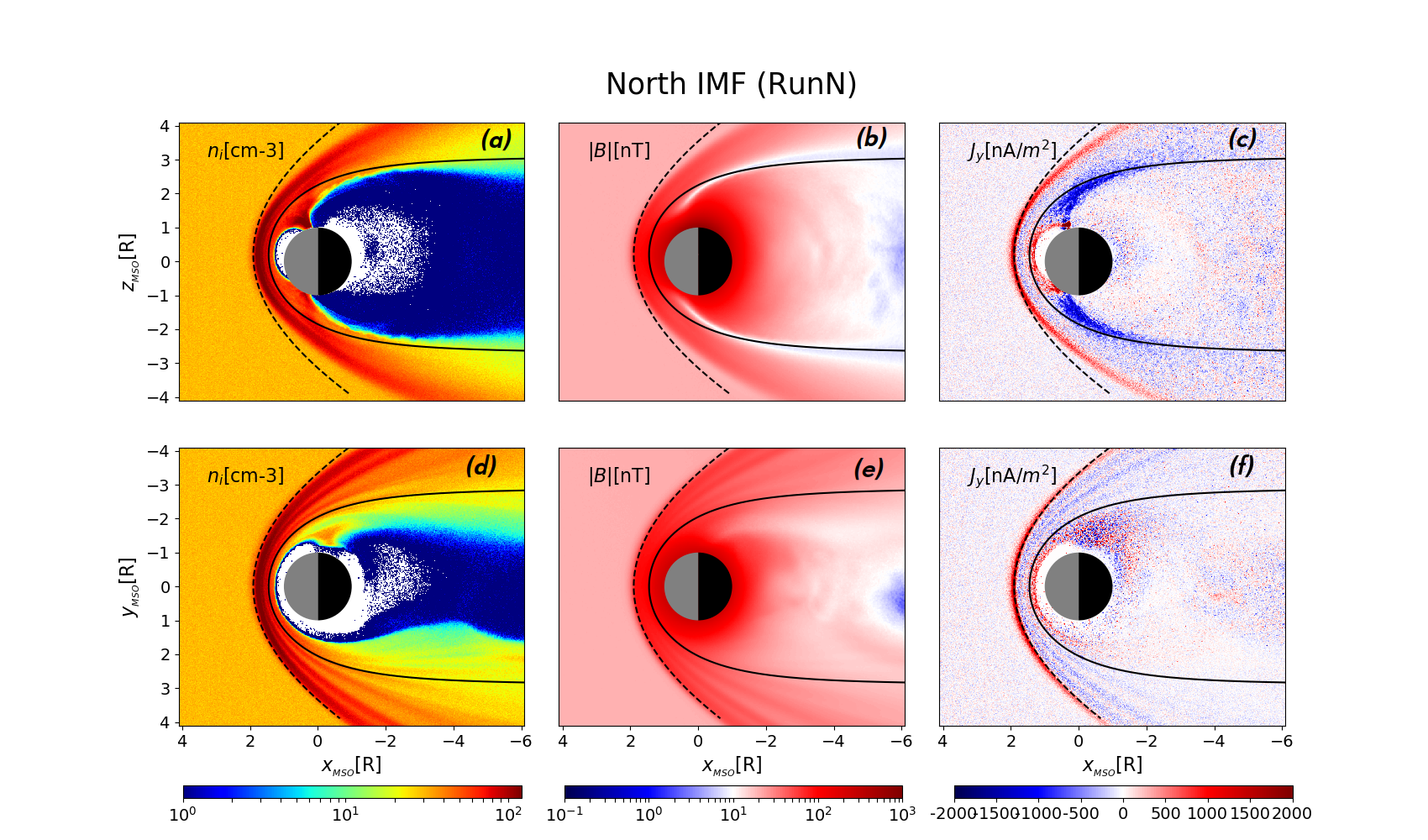}
    \includegraphics[width=\linewidth]{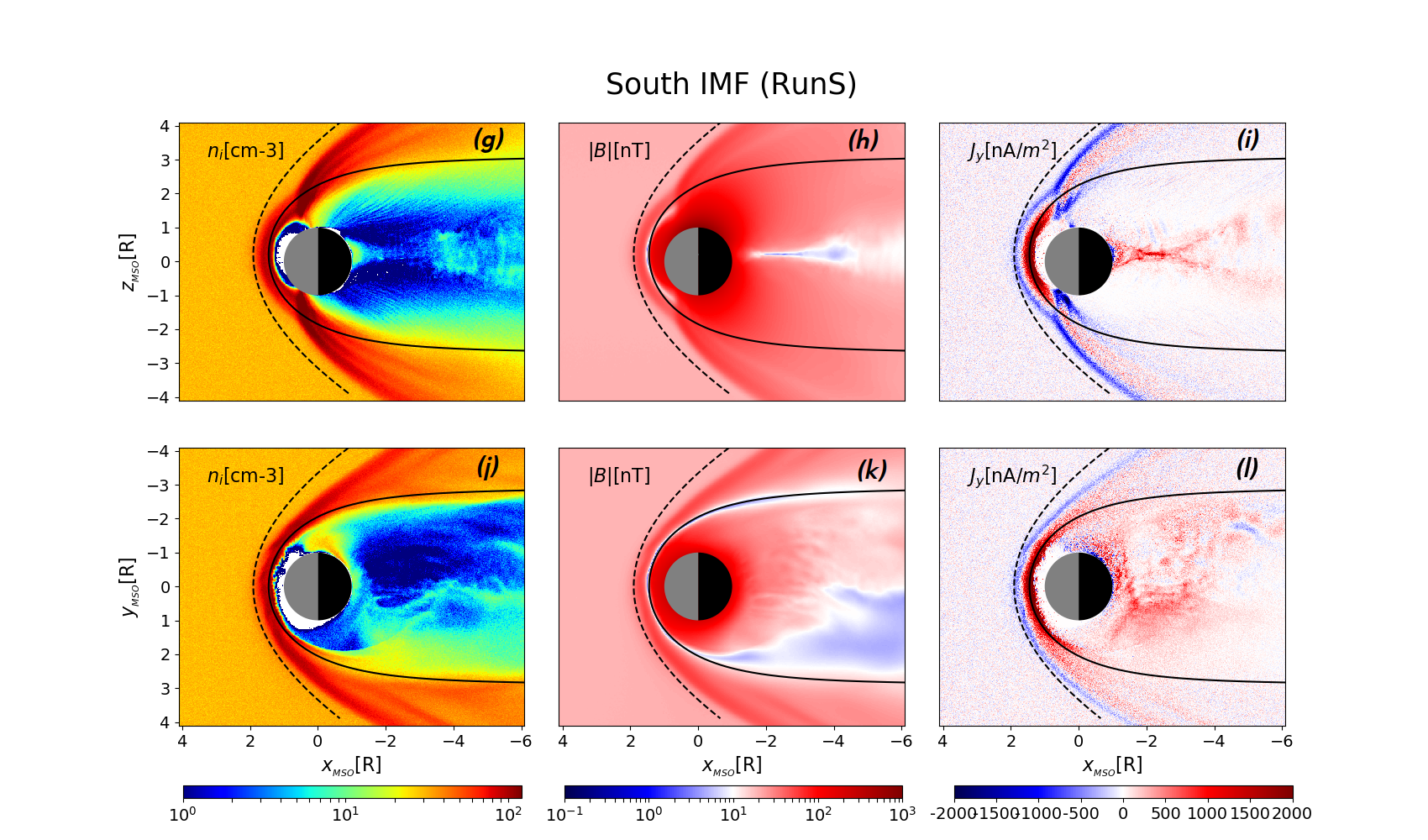}
    \caption{Overview of the large scale structure and magnetosphere boundaries in our simulations. Panels (a)-(f) show the results of RunN. Panels (a)-(c) show dipolar $Y_{_{MSO}}$ = 0 cuts of the ion density $n_i$ (a), magnetic field amplitude $|\bf{B}|$ (b) and plasma current density along the y-axis $J_y$ (c). Panels (d)-(f) show equatorial $Z_{_{MSO}}$ = 0 cuts of the same quantities for RunN.
    Panels (g)-(l) show the results of RunS in the same format as those of RunN.
    All plots are done at time $t$ = 15.9 $R/V_x$. 
    The average bow shock (dashed black line) and magnetopause (solid black line) profiles found by~\citet{Winslow2013} using MESSENGER observations are added. 
    The white region around the planet corresponds to grid cells with zero macro-particles.}
    \label{fig:fig1-nBJ}
\end{figure*}
\begin{figure*}
    \centering
    \includegraphics[width=\linewidth]{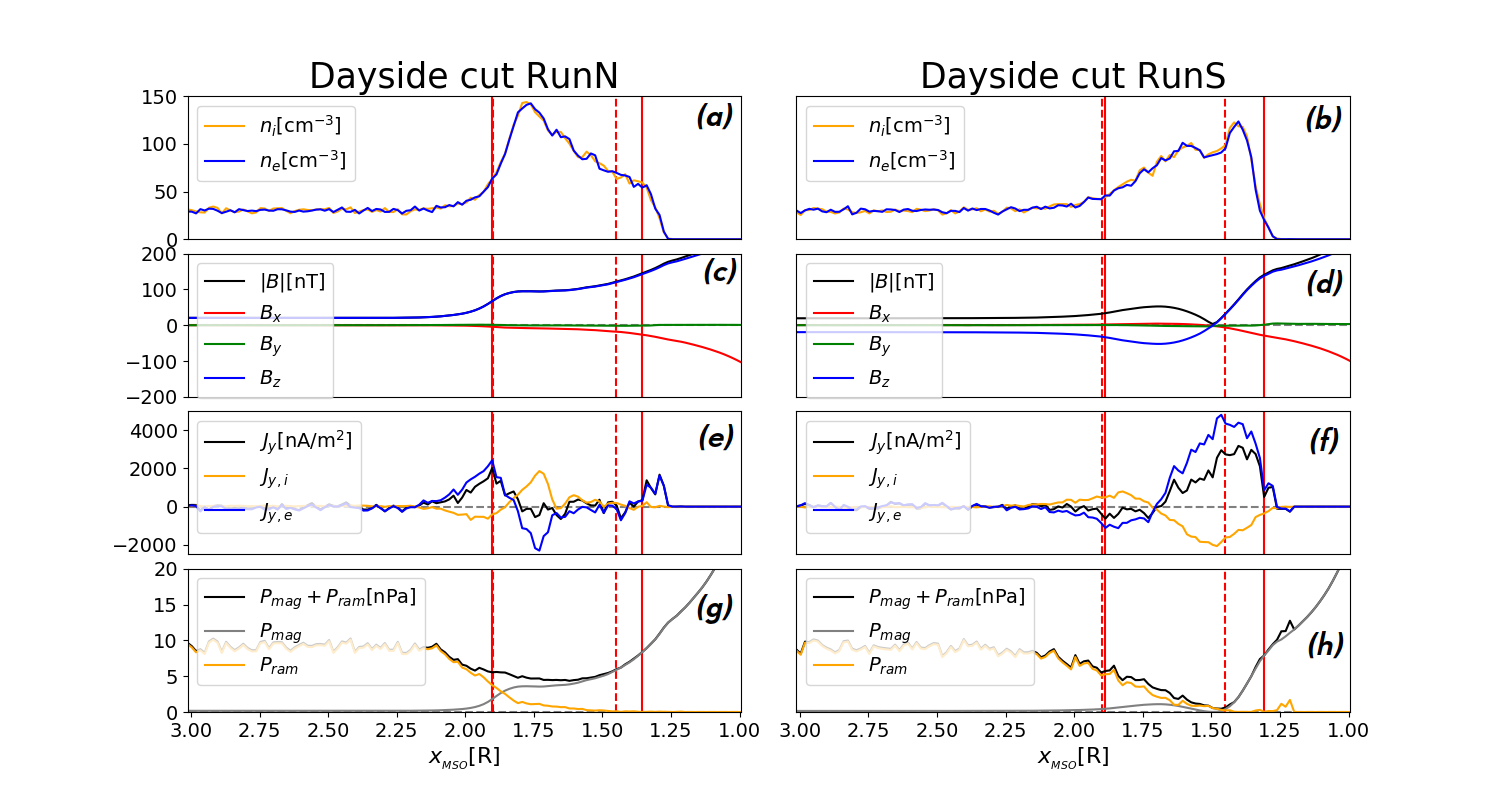}
    \caption{Cut of the results of our two simulations RunN (left panels) and RunS (right panels) along the subsolar line (x-axis).  
    Panels (a)-(b): proton and electron densities.
    Panels (c)-(d): magnetic field components and amplitude.
    Panels (e)-(f): current density along $Y_{_{MSO}}$ of protons, electrons and total.
    Panels (g)-(h): magnetic $P_{mag} \equiv B^2/2\mu_0$, ram $P_{ram} \equiv m_in_iV^2_{x,i}+m_en_eV^2_{x,e}$ and total pressures. 
    These cuts are performed at the same time as Fig.~\ref{fig:fig1-nBJ} ($t$ = 15.9 $R/V_x$).
    We highlight the bow shock and magnetopause stand-off distances obtained from our model (solid red vertical lines) and the ones obtained using the average model of~\citet{Winslow2013} constrained by MESSENGER observations (dashed red vertical lines).}
    \label{fig:fig2-cut-dayside}
\end{figure*}


\subsection{Validation of the global fully-kinetic model}

We start by validating our model by showing that the large scale structure of the system typically obtained by fluid and hybrid models is correctly reproduced, as shown in Fig.~\ref{fig:fig1-nBJ}.

Indeed, both simulations show the formation of a bow shock standing in front of the planet, the magnetosheath with increased density and magnetic field amplitude, the magnetopause with its strong current $J_y$, and the magnetosphere cavity with very low density and high magnetic field. Such large scale configuration is shown in Fig.~\ref{fig:fig1-nBJ} for both simulations using the ion density $n_i$ [cm$^{-3}$] (left panels), the magnetic field amplitude $|\bf{B}|$ [nT] (central panels) and the plasma current $J_y$ [nA/m$^2$] (right panels). In Fig.~\ref{fig:fig2-cut-dayside} we show a cut along the subsolar line $Y_{_{MSO}}$ = $Z_{_{MSO}}$ = 0.

The observed bow shock and magnetopause are compared to their nominal shape and position obtained from MESSENGER data~\citep{Winslow2013}. The authors modeled the bow shock using a paraboloid model (with parameters $X_0$ = 0.5 $R$, $e$ = 1.04, $p$ = 2.75 $R$;~\citealt{Slavin2009}) and the magnetopause using the~\citet{Shue1997} model (with parameters $R_{ss}$ = 1.45 $R$, $\alpha$ = 0.5).
The resulting bow shock and magnetopause profiles are shown in Fig.~\ref{fig:fig1-nBJ} using dashed and solid black lines, respectively. 
Overall, the bow shock and magnetopause found in our simulations are in good agreement with their nominal shapes. As expected, a better agreement is observed in the equatorial plane (see Fig.~\ref{fig:fig1-nBJ}), whereas some discrepancy is found in the dipolar plane due to the broken assumption of cylindrical symmetry assumed in these relatively simple analytical models (the magnetic field along the z-axis breaks the symmetry).

We compute the bow shock and magnetopause stand-off distances as the maximum of the current density $|\textbf{J}(x)|$ and the point where the magnetic pressure equals the solar wind ram pressure, respectively. The stand-off positions of the bow-shock and magnetopause obtained from our simulations are in agreement with the nominal values reported by~\citet{Winslow2013}, as shown in Fig.~\ref{fig:fig2-cut-dayside}. The relative discrepancy for the bow shock and the magnetopause is of the order of $1\%$ and $10\%$, respectively. 
Interestingly, the discrepancy for the magnetopause is larger in both runs when compared to the one for the bow shock. Nonetheless, similar discrepancies of the order of 10\% have been found by~\citet{Aizawa2021} using global hybrid and fluid models.
We conclude that our fully-kinetic model satisfactorily reproduces the large scale structures of the Hermean magnetosphere. 
This validates our model at large-scales and paves the way for further analysis of the small-scale electron physics in the next sections.

\subsection{Impact of electron physics at large scales: boosting the charge current at the boundaries}

Past works using hybrid or fluid models found a total current density at the bowshock and magnetopause on the order of 100 nA/m$^2$~\citep{Janhunen2004,Benna2010,Exner2020,Aizawa2021}.
In our fully-kinetic simulations instead, we find significantly higher values of the current density on the order of 1000 nA/m$^2$, as shown in Fig.~\ref{fig:fig2-cut-dayside}e-f.
Such higher values are due to the presence of a dominant electron current at the magnetic field boundaries, as shown in Fig.~\ref{fig:fig2-cut-dayside}e-f.

Past global fluid~\citep{Benna2010,Aizawa2021} and hybrid~\citep{Janhunen2004,Exner2020,Aizawa2021} simulations treated electrons as a mass-less neutralizing fluid. Under such assumptions, the electron current was computed from Amp\`ere's equation using the curl of the magnetic field and the proton current. Differently from those works, in our model, the electron current is computed self-consistently from their distribution function evolving under Vlasov-Maxwell system of equations.
Since the magnetic field amplitude on both sides of the magnetospheric layers in our work is consistent with past simulations, we conclude that the width of these layers (bow shock and magnetopause) tends to be overestimated in global fluid and hybrid models with mass-less electrons. This result highlights the key role of electrons as current carrier in the magnetic boundaries of the Hermean magnetosphere, and the importance of an appropriate modeling of this species even in global simulations to satisfactorily address large-scale planetary boundaries. In other words, even if the location of the magnetospheric boundaries is well-reproduced with models using fluid mass-less electrons, a more physically correct treatment of the electrons is required to draw conclusions on the physics at the magnetopause at and below the ion scale.

\subsection{IMF configuration and magnetic reconnection: feedback from small to large scales}

By changing the direction of the IMF, a strong reconfiguration of the global structure of the magnetosphere is observed, as shown in Figs.~\ref{fig:fig1-nBJ}-\ref{fig:fig2-cut-dayside} where we show the results of the two different IMF configurations. Such global reconfiguration is primarily driven by magnetic reconnection at the nose (around $X_{_{MSO}}\approx$ 1.5 $R$) and in the tail (around $X_{_{MSO}}\approx$ -2 $R$) of the magnetosphere. It is much more active in RunS (run with southward IMF) as compared to RunN (run with northward IMF). These regions are highlighted in Figs.~\ref{fig:fig1-nBJ}i-\ref{fig:fig2-cut-dayside}f for RunS by the plot of the current $J_y$. The position of the neutral line in the tail in RunS is in good agreement with MESSENGER observations~\citep{Poh2017}.

Magnetic reconnection has a strong impact on the global dynamics of the magnetosphere since it rapidly changes the large-scale magnetic field topology thus allowing for an efficient plasma injection in the magnetosphere and fast conversion of magnetic to particle's energy~\citep{Vasyliunas1975,Yamada2010,Treumann2013}. Magnetic reconnection is intrinsically a multi-scale phenomenon strongly coupling the ion and sub-ion kinetic scale dynamics. Therefore, a precise description of the physics at the electron scale is important to correctly capture the features observed in laboratory and space plasmas. This is usually achieved by means of fluid or hybrid models with inertial electron closures~\citep{Wang2015,Ng2017,Finelli2021,Fadanelli2021,Jain2022} or by means of fully-kinetic models~\citep{Pritchett2001,Pritchett2001_GEM,Divin2007,Haggerty2015} as in this work.

Magnetic reconnection drives electron acceleration and heating thus feeding the global magnetosphere circulation pattern with supra-thermal electrons. We focus on such strong impact of magnetic reconnection on the electron dynamics in the next section.

\section{Electron heating and circulation in the magnetosphere: global feedback of magnetic reconnection}\label{sec:Result_electrons}

\begin{figure*}
    \centering
    \includegraphics[width=\linewidth]{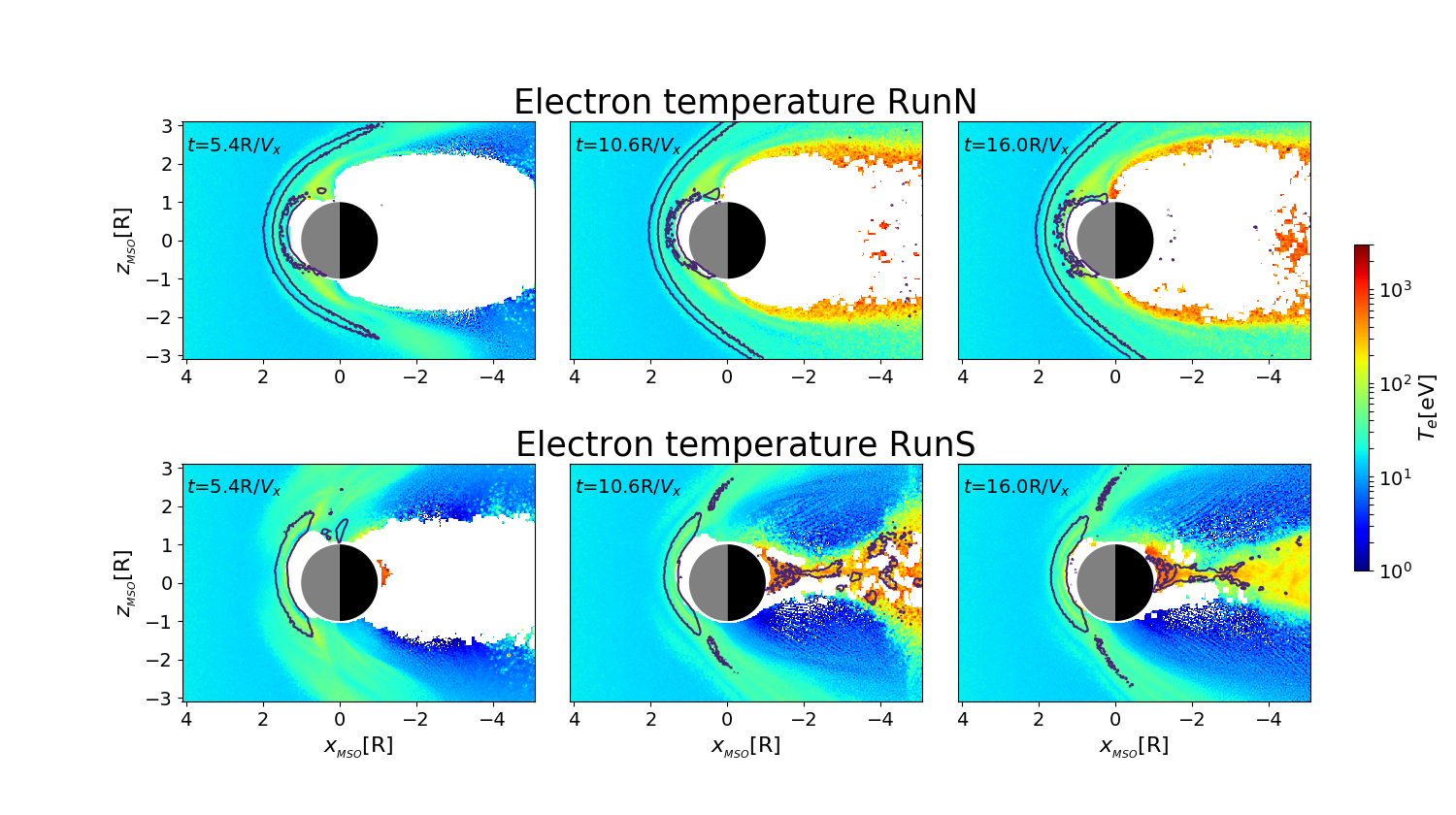}
    \caption{Time evolution of the electron temperature in the nose and tail reconnection regions. Dipolar $Y_{_{MSO}}$ = 0 cut of simulations RunN (top panels) and RunS (bottom panels) at different times (time advancing from left to right). We show the colormap of the electron temperature $T_e=(2T_{\bot,e}+T_{\parallel,e})/3$ in units of [eV] in logarithmic scale, superposed to the isocountour of the out-of-plane current $J_y$ = 300 nA/m$^2$ (black solid lines). 
    The white region around the planet corresponds to grid cells with a number of macro-particles too small to reliably compute the value of $T_e$.}
    \label{fig:fig3-electrons-temp}
\end{figure*}
\begin{figure*}
	\centering
	\includegraphics[width=\linewidth]{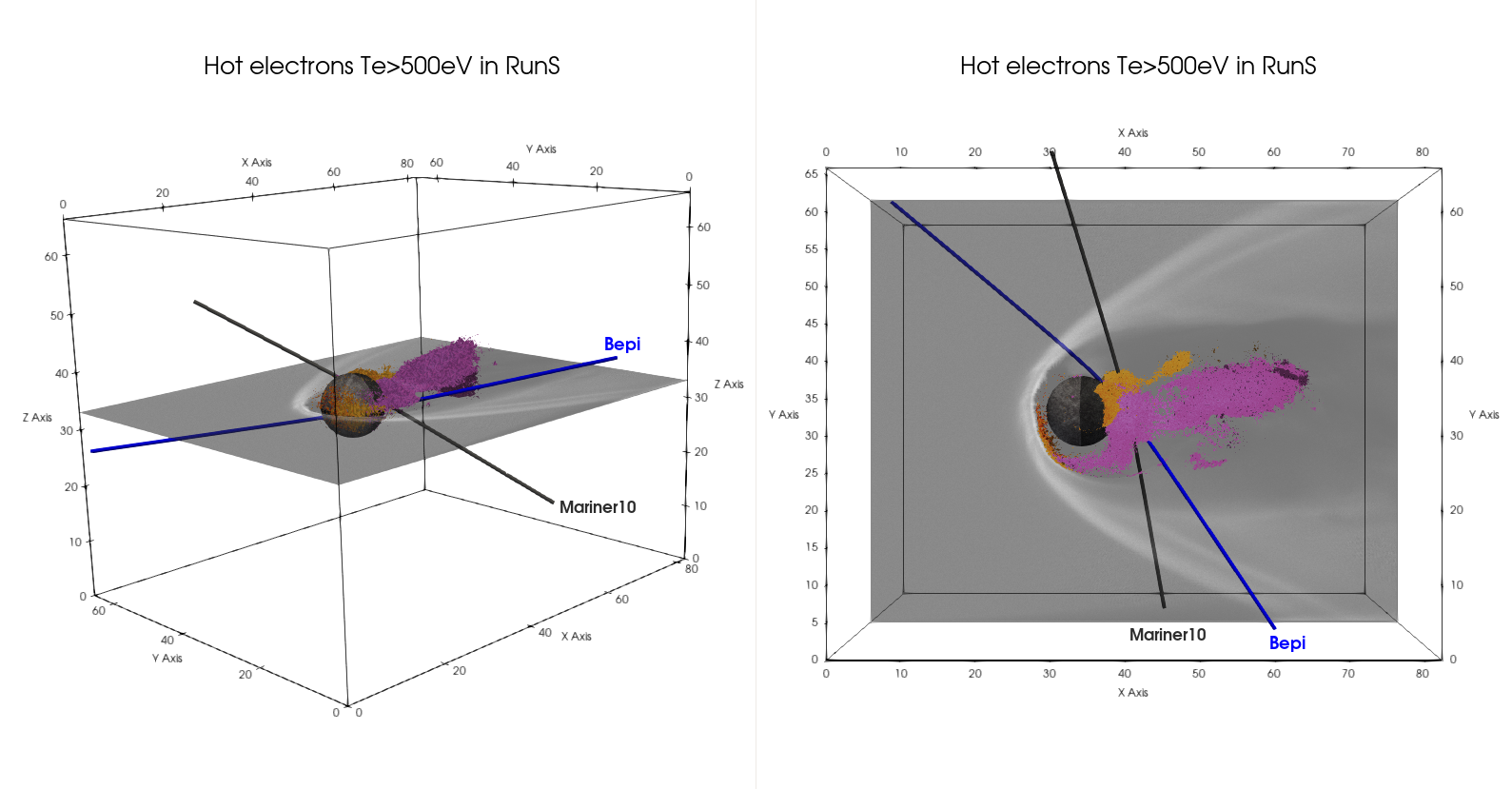}
	\caption{Three dimensional view of the high-energy electron population in simulation RunS. Grid cells with values of perpendicular (parallel) temperature greater than $500$ eV are shown in orange (purple). We also show an equatorial cut of the electron density in greyscale. 
	Both quantities are computed at time 15.9 $R/V_x$, same as Figs.~\ref{fig:fig1-nBJ}-\ref{fig:fig2-cut-dayside}. 
	Mariner10 and BepiColombo first Mercury flyby trajectories are shown using grey and blue solid lines, respectively. In this figure, axis are not in MSO coordinates.} 
	\label{fig:fig4-3dview-Te}
\end{figure*}
\begin{figure*}
	\centering
	\includegraphics[width=\linewidth]{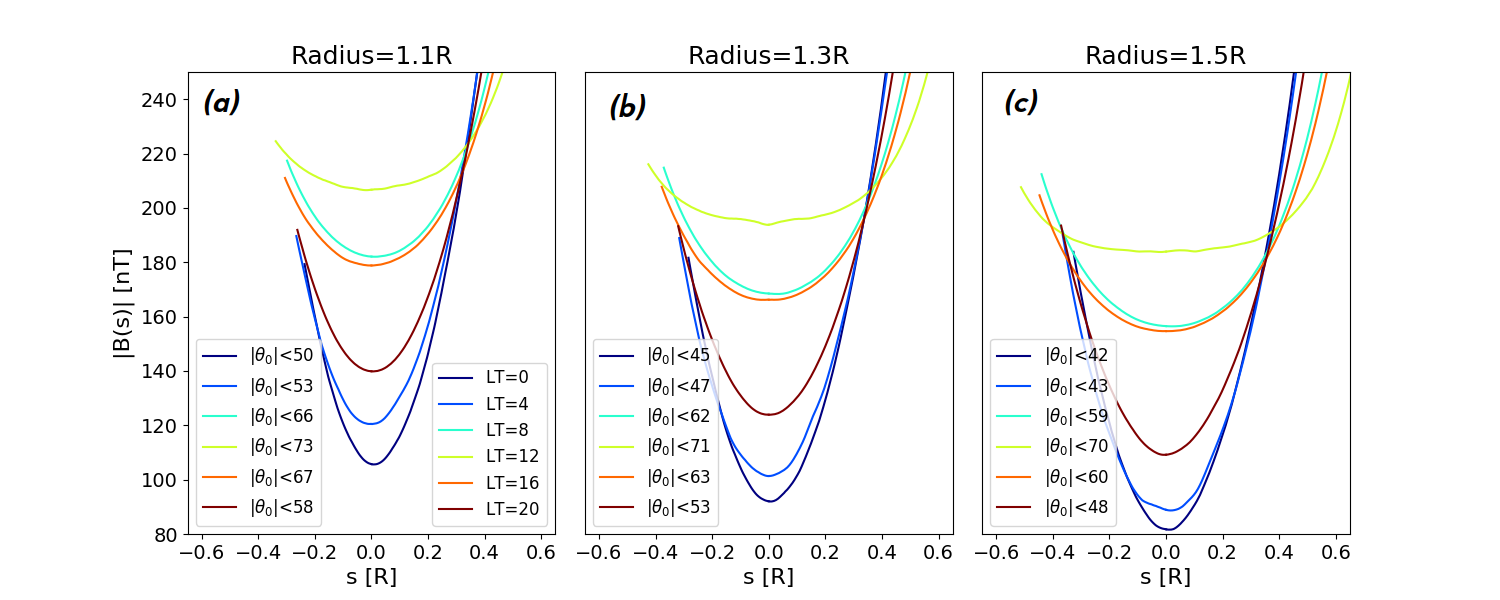}
	\caption{Magnetic field amplitude along magnetic field lines $|B(s)|$ [nT] computed at different local times (LT), LT 12 corresponds to subsolar longitude (yellow) and LT 0 corresponds to local midnight (dark blue). Panels (a)-(c) correspond to different equatorial distances from the center of the planet: 1.1 $R$ (a), 1.3 $R$ (b) and 1.5 $R$ (c). Resulting values for the loss-cone angle from Eq.~\ref{eq:loss-cone_adiabat} are also shown in each panel. 
	Plots done at time 15.9 $R/V_x$, same as Figs.~\ref{fig:fig1-nBJ}-\ref{fig:fig4-3dview-Te}.} 
	\label{fig:fig5-Blines}
\end{figure*}

As outlined in Sec.~\ref{sec:Introduction}, the role of the electrons in the interaction between the solar wind and the Hermean magnetosphere is still a matter of debate. In this section, we show that electrons are strongly energized and partially trapped in the Hermean magnetosphere under southward IMF conditions.
We focus on the results of RunS since electron heating and injection in the magnetosphere are much more efficient for southward IMF conditions, as shown in Fig.~\ref{fig:fig3-electrons-temp} by comparing the two runs. In the case of northward IMF -- although electrons are heated efficiently at the magnetopause boundary, their flux across the magnetopause is negligible, in turn virtually no electrons interact directly with the planetary dipole. A more detailed comparison between the two runs is presented in Sec.~\ref{sec:Result_comparison-data}.

The electron dynamics in RunS is mainly driven by (i) the occurrence of magnetic reconnection causing strong acceleration and heating and (ii) the planetary dipole magnetic field causing trapping and drifting of electrons inside the magnetosphere. In the following, we discuss these two mechanisms separately. 

First, magnetic reconnection affects plasma parcels flowing through the reconnection region by (i) accelerating the plasma (increase of the plasma bulk flow velocity) up to the Alfv\'en speed $V_{A,in}$ computed from the inflow plasma parameters\footnote{In the symmetric case, the inflow Alfv\'en speed is $V_{A,in}=B_r/\sqrt{4\pi m_i n}$, where $B_r$ is the reconnecting magnetic field. While, in the asymmetric case the generalization of this expression is $V_{A,in}=\sqrt{B_1 B_2/4\pi m_i} (B_1+B_2)/(n_1B_1+n_2B_2)$, as reported in~\citet{Cassak2007}.} and (ii) heating the plasma (increase of the thermal speed of the particles) by an amount proportional to the inflow magnetic energy $\Delta T \approx 0.1 m_i V^2_{A,in}$~\citep{Phan2014,Shay2014,Haggerty2015}. All in all, the magnetic field energy of the plasma in the inflow acts as an energy reservoir for the kinetic energy in the outflow.

In RunS, the two reconnection sites at the nose and tail of the magnetosphere present different electron temperatures in the outflows. In Fig.~\ref{fig:fig3-electrons-temp} by showing the evolution of the electron temperature $T_e=(T_{e,\parallel}+2T_{e,\bot})/3$ in the dipolar $Y_{_{MSO}}$ = 0 plane (the reconnection plane) we highlight the circulation of solar wind electrons from the nose of the magnetopause to the tail finally entering the magnetosphere cavity. The electron temperature increases with respect to its upstream value $T_{e,_{SW}}\approx$20 eV up to 100-200 eV in the nose, reaching in the tail larger values of the order of few keVs. This asymmetry in electron heating between nose and tail reconnection regions is due to the different inflow plasma parameters: while in the nose the magnetic energy is $m_iV^2_{A,in}\approx$1 keV, in the tail $m_i V^2_{A,in}\approx$10 keV, given the lower density and higher magnetic field amplitude in the lobes as compared to the magnetosheath. Thus, more magnetic energy is available for conversion into kinetic energy in the tail as compared to the nose. This simple estimate explains why we observe electrons with higher energy in the tail (by around a factor 10) with respect to the nose of the magnetosphere.

Magnetic reconnection allows for the injection of solar wind electrons into the Hermean magnetosphere. As shown by the time evolution in Fig.~\ref{fig:fig3-electrons-temp}, from time $t\approx10 R/V_x$ the nightside part of the magnetosphere starts to be populated with high energy electrons ejected from the reconnection site in the tail around $X_{_{MSO}}\approx -2R$ (such planetward flow of plasma is also called substorm, in analogy with Earth; \citealt{Christon1987}).
Once inside the Hermean magnetosphere, the motion of such electron's substorm is driven by the dipolar structure of the magnetic field. Similarly to Earth, in this region electrons bounce back-and-forth along closed magnetic field lines with period $\tau_{B}$. Electrons also move around the planet following a longitudinal drift motion with period $\tau_{i}$ due to curvature drift. These motions have very different time scale $\tau_{ce} \ll \tau_{B} \ll \tau_{i}$ (here $\tau_{ce}$ is the electron gyro-period as defined in Sec.~\ref{sec:Model}), thus allowing us to treat them separately using the adiabatic approximation~\citep{Northrop1963}. 

Using the adiabatic approximation, the bouncing motion of electrons along a given magnetic field line is described by the conservation of energy:
\begin{equation}\label{eq:energy_adiabat}
	E = \frac{1}{2}m_e v^2_{\parallel}(s) + \mu B(s) 
\end{equation}
where $\mu=m<v^2_\bot>/2B$ is the first adiabatic invariant. Here brackets $<>$ indicate average over the gyromotion and $s$ is the curvilinear coordinate along magnetic field lines. Similarly to the problem of an harmonic oscillator, the inversion points $s_m$ are defined as those where the kinetic energy goes to zero, thus:
\begin{equation}\label{eq:Bm_adiabat}
	B(s_m)= \frac{E}{\mu} = \frac{B(s_0)}{\sin^2 \theta_0}
\end{equation}
where $\theta_0$ is the initial particle pitch angle, $\theta_0 \equiv \tan^{-1} (v_{0,\bot} / v_{0,\parallel})$, computed at $s_0$ (the point where the magnetic field line crosses the magnetic equator). From Eq.~\ref{eq:Bm_adiabat} and using the value of the magnetic field amplitude at Mercury's southern pole $B_{south}$ (of the order of 200 nT), we get the following loss-cone equation for electrons:
\begin{equation}\label{eq:loss-cone_adiabat}
|\theta_0| < \sin^{-1} \sqrt{ \frac{B(s_0)}{B_{south}} }
\end{equation}
This equation defines a family of loss-cones associated to different magnetic field lines (parameterized by $s_0$) varying with radius and local time (hereafter LT). Thus, spatial variations of the magnetic field amplitude $B(s_0)$ in the equatorial plane induce variations in the electron pitch-angle distribution around the planet.

In RunS, we observe such a variation of electron distribution function around the planet as a function of local time. This is shown in Fig.~\ref{fig:fig4-3dview-Te} using three-dimensional (3D) spatial distributions of electrons with $T_{e,\bot}>500$ eV (orange) and $T_{e,\parallel}>500$ eV (purple). Less energetic electrons are observed in the dayside sector compared to the nightside, and the few observed there have pitch angles close to $90^{\circ}$. This means that most of the electrons do not complete a full-drift orbit around the planet creating a stable continuous belt as on Earth, but rather bounce along magnetic field lines in the nightside before drifting dawn-ward and eventually falling on the planet surface. In the following, we refer to this partial circulation pattern as nightside-trapping.

To investigate such peculiar property of the Hermean environment, we look at the variation of the profiles $B(s)$ with local time (LT) and radial distance, as shown in Fig.~\ref{fig:fig5-Blines}.
The compression of the dipole magnetic field by the solar wind stretches the field lines at nightside (blue-black lines in Fig.~\ref{fig:fig5-Blines}) and compresses the field lines at dayside (yellow-orange lines in Fig.~\ref{fig:fig5-Blines}). Such a compression  at dayside increases the magnetic field amplitude $B(s_0)$ at the subsolar equator, which in turn increases the loss-cone angle for electrons according to Eq.~\ref{eq:loss-cone_adiabat}.
The loss-cone angles resulting from Eq.~\ref{eq:loss-cone_adiabat} are reported in Fig.~\ref{fig:fig5-Blines}a-c for varying LT (0 corresponds to local midnight and 12 corresponds to local noon) and radial distance (from 1.1 to 1.5 $R$).
For any radial distance close-enough to the planet, the loss-cone angle steadily increases by moving from nightside (angle of the order of $45^{\circ}$) to dayside (angle of the order of $70^{\circ}$). This explains the nature of the electron nightside-trapping observed in RunS. On top of that, the values of the loss-cone angles reported in Fig.~\ref{fig:fig5-Blines} point out the higher probability of having trapped particles in the outer shells ($\approx$ 1.5 R) rather than very close to the planet ($\approx$ 1.1 R). This interpretation of adiabatic nightside-trapping is consistent with the spatial distribution of high-energy electrons around the planet shown in Fig.~\ref{fig:fig4-3dview-Te}.

The loss-cone mechanism creates an inhomogenous distribution of high-energy electrons inside the Hermean magnetosphere (in the range $R\approx$ 1.1 - 1.5) with anisotropic energy distribution function ($T_{e,\bot}>T_{e,\parallel}$).
Such electron population is expected to be the target of observation by satellites, able to measure \textit{in situ} electron energy distribution functions (hereafter eEDF).
In the next section, we will address this point by showing in our simulations synthetic eEDF sampled along Mariner10 and BepiColombo first Mercury flyby trajectories.

\section{Comparison between observations and simulations}\label{sec:Result_comparison-data}

\begin{figure*}
	\centering
	\includegraphics[width=\linewidth]{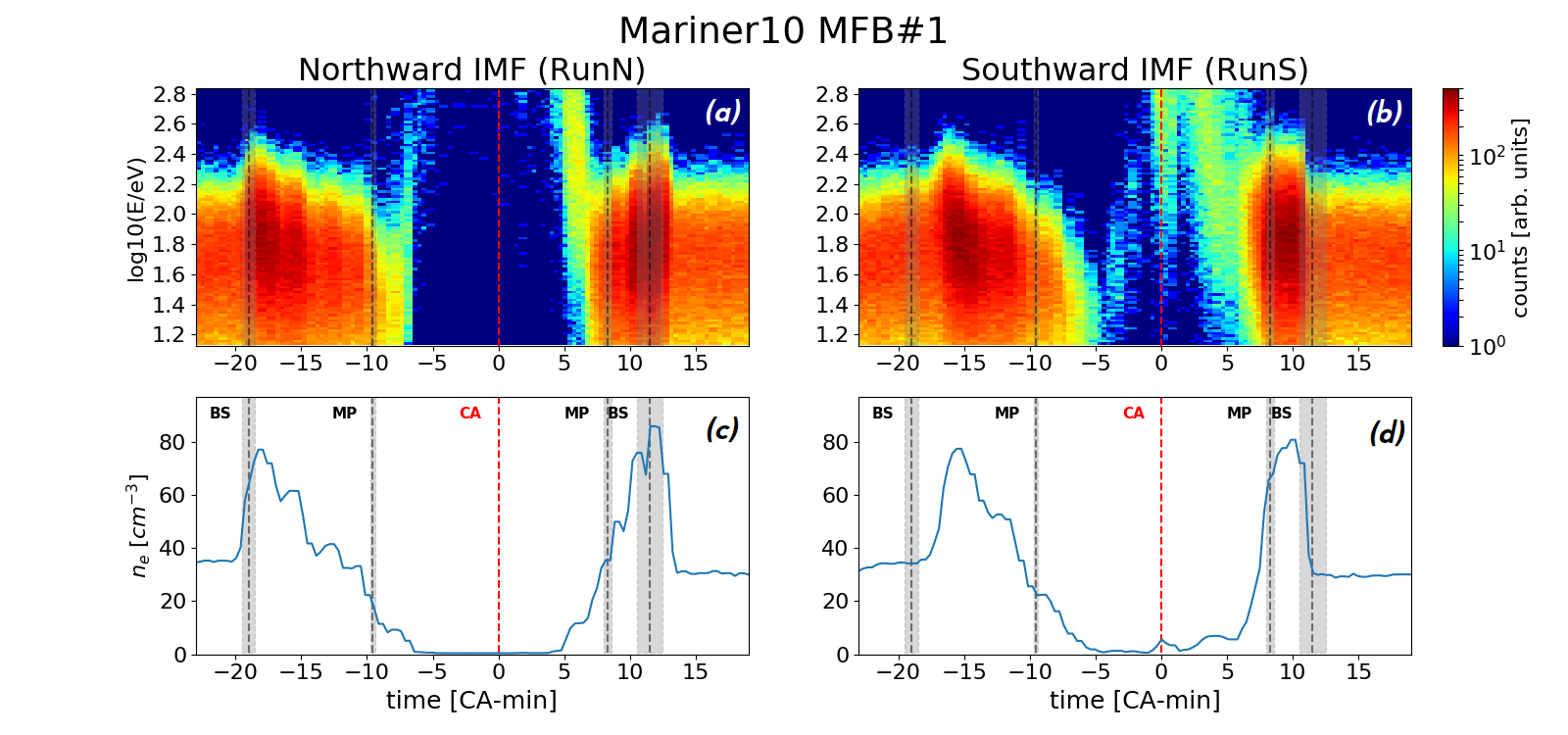}
	\includegraphics[width=\linewidth]{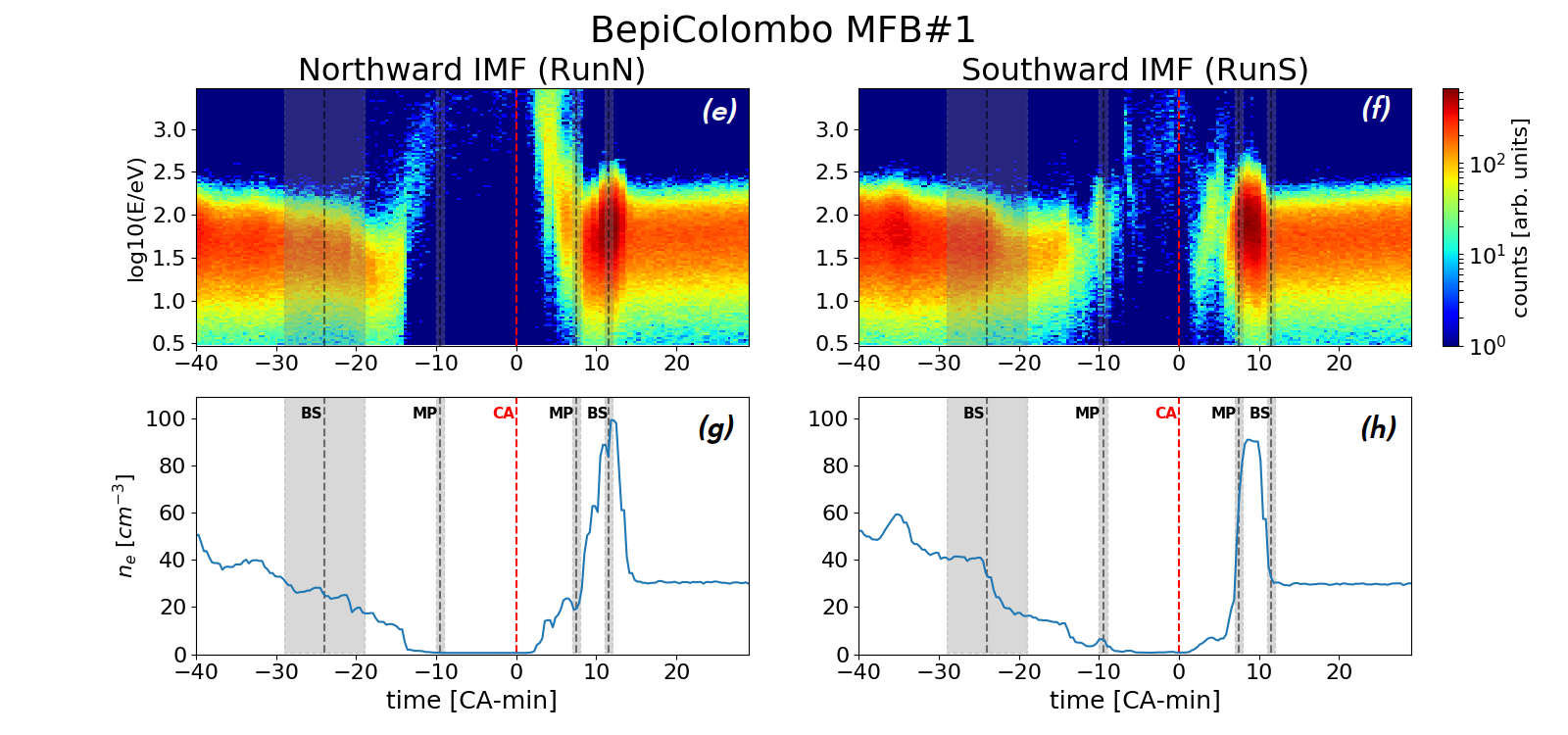}
	\caption{Synthetic electron energy distribution functions (eEDF) sampled along Mariner10 (a)-(d) and BepiColombo (e)-(h) first Mercury flybys in our simulations boxes using northward IMF (left panels) and southward IMF (right panels) conditions.
	The plots are done using simulation results at time=14.7$R/V_x$.
	The energy ranges and resolutions used to compute the eEDF are the same as those of the instruments PLS (onboard Mariner, top) and MEA1 (onboard BepiColombo, bottom). 
	The horizontal axis shows the time from closest approach (CA) in minutes, $t=0$ is highlighted using vertical red dashed line.
	On the time axis, we highlight the bow shock (BS) and magnetopause (MP) crossings obtained by the two spacecraft \textit{in-situ}, these are highlighted using grey areas and vertical black dashed lines.
	The electron density is obtained by integrating the synthetic eEDF in energy at each given time.} 
	\label{fig:fig6-observations}
\end{figure*}

The simulation results discussed so far are of particular interest for the interpretation of~\textit{in situ} electron observations at Mercury. To ease comparison with observations, we sample synthetic electron energy distribution function (eEDF) from our two simulations along Mariner10 and BepiColombo first-Mercury-flyby trajectories, as shown in Fig.~\ref{fig:fig6-observations}.

 To compute the energy spectra in Fig.~\ref{fig:fig6-observations} from our simulations, we select the macro-particles lying in the neighborhood of the spacecraft trajectory. In particular, we use a range of $\pm4$ cells (equal to $\approx \pm$0.06R) in all directions in order to minimize the particle noise and mimic the effect of particle precipitation on the instruments in an acquisition time $\delta t$. Moreover, we assume that the field of view of the instruments spans the whole $4\pi$ solid angle in an acquisition time $\delta t$. Although this assumption is usually partially verified by the instruments, the main features of the spectra in Fig.~\ref{fig:fig6-observations} are expected to hold also in the case of limited field of view.
 To ease the comparison with \textit{in situ} observations, the spectra in Fig.~\ref{fig:fig6-observations}a-b have the same energy range and resolution of the Mariner10/PLS instrument~\citep{Ogilvie1974} and those in Fig.~\ref{fig:fig6-observations}e-f of the BepiColombo/MEA1~\citep{Saito2021} instrument operating in solar-wind mode during the flyby. Moreover, in Fig.~\ref{fig:fig6-observations} we report the bow shock and magnetopause crossing times obtained by the two spacecraft \textit{in situ} at Mercury as discussed in~\citet{Russell1988} (panels a-d) for Mariner10 and in~\citet{Andre2022} for BepiColombo (panels e-h).
 In the following, we discuss these two spacecraft flybys modeling results separately.

Mariner10 synthetic eEDF and electron densities obtained from our simulations, shown in Fig.~\ref{fig:fig6-observations}a-d, present clear signatures of (i) inbound bow shock crossing at time $t\approx[-20,-18]$, (ii) inbound magnetopause crossing at time $t\approx-7$, (iii) outbound magnetopause crossing at time $t\approx[6,7]$, and (iv) outbound bow shock crossing at time $t\approx[12,14]$ (times are given in minutes to closest approach).
These crossing times are consistent with those observed by Mariner10~\citep{Russell1988}, the latter are and shown in Fig.~\ref{fig:fig6-observations} using vertical black dashed lines and grey areas. A more precise comparison of the crossing times between observations and simulations would require an \textit{ad-hoc} simulation initialized with the upstream solar wind parameters observed by Mariner10. Such further step will be addressed in future works. Here we mainly focus on the qualitative features of the eEDF in the Hermean magnetosphere. Indeed, inside the magnetosphere ($t\approx[-7,+7]$) in both simulations the plasma density is strongly depleted and we observe different signatures in the two runs:
\begin{itemize}
    \item In RunN electrons with energies up to few keV are observed around the magnetopause (the mechanism accelerating these electrons however remains unclear) and no electrons are observed inside the magnetosphere.
    \item In RunS electrons with energies up to tens of keV (well above the instrumental cutoff of 687eV) are encountered around and after closest approach ($t\approx[0,+7]$). These high energy electrons are produced by magnetic reconnection in the tail as discussed in Sec.~\ref{sec:Result_electrons}.
\end{itemize}
This result support the interpretation of Mariner10/PLS observations outlined in past works~\citep{Ogilvie1974,Christon1987}, stating that Mariner10 observed substorm injected electrons around closest approach with energies well above the instrumental cutoff, and can be further used to interpret the recent observations by BepiColombo at Mercury.

BepiColombo synthetic eEDF and electron densities obtained from our simulations, shown in Fig.~\ref{fig:fig6-observations}e-h, present similar signatures to those of Mariner10. However two main differences are observed.
First, for BepiColombo the inbound bow shock crossing is less sharp than for Mariner10. This is due to the large distance from the planet of the spacecraft at the encounter of the inbound bow shock (see the trajectory in Fig.~\ref{fig:fig4-3dview-Te}). BepiColombo first flyby trajectory makes plasma observations more asymmetric between inbound and outbound as compared to Mariner10.
Second, signatures of high energy electrons around closest approach are fainter for BepiColombo than for Mariner10. This effect can be understood again in terms of trajectories. From Fig.~\ref{fig:fig4-3dview-Te}, we note that BepiColombo passes beneath the high energy region in the nightside while Mariner10 passes right through such region (shown as purple/orange in Fig.~\ref{fig:fig4-3dview-Te}).

We stress that a more precise, quantitative comparison between observations and simulations would require a good knowledge of the upstream solar wind parameters which define the precise values of, among others, the time of the bow shock and magnetopause crossings, the energy of the electrons inside the magnetosphere, and the density of electrons in the magnetosheath. Thus, the results of this section are limited to the qualitative main features observed by Mariner10 and BepiColombo first Mercury flybys.

A comparison of the synthetic eEDF between RunN and RunS demonstrate that by changing only one parameter (the IMF direction) a sudden change in the energy and density of electrons inside the magnetosphere is induced, as shown in Fig.~\ref{fig:fig6-observations}. We suggest that the Hermean environment responds to such changes in the solar wind by a corresponding reconfiguration as from RunN to RunS or viceversa. The characteristic time scale for such a reconfiguration is expected to be comparable to the Dungey cycle $\tau\approx 10 R/V_x\approx$ minutes. Thus, it is only in the case of stable solar wind conditions on time scale longer than a few minutes that observational signatures in the eEDF inside the magnetosphere can be linked to the upstream IMF direction.

\section{Discussion and conclusions}\label{sec:Sec4}

This work represents a first step towards global fully-kinetic modeling of planetary magnetospheres in which the electron dynamics is included self-consistently from the global planetary scale down to the gyro-radius scale. The high computational needs of this model impose strong constraints on the possible scale separation between planet, ion and electron scale length, as discussed at the end of Sec.~\ref{sec:Model}.
Scaling down the ion-to-electron mass ratio and the plasma-to-cyclotron frequency ratio remains a classic procedure in fully-kinetic simulations, which effects are extensively discussed in past works~\citep{Bret2010,Le2013,Lavorenti2021}.
On the other hand, using a smaller planet radius $\approx$ 230 km instead of the realistic Mercury radius $\approx$ 2400 km may have a significant impact on simulation's results (in our work here the scale separation between planet radius and ion gyro-radius remains of about a factor 10).

First, scaling down the planet radius can have an impact on the global shape of the magnetosphere. To avoid this problem, the magnetic moment of the planet is scaled down proportionally to $R^3$ in order to preserve the shape of the magnetosphere. Therefore the global shape of the interaction region is preserved, and in particular the bowshock $d_{BS}/R$ and magnetopause $d_{MP}/R$ stand-off distance in units of the planet radius are preserved, as discussed in Sec.~\ref{sec:Results_large}.
Moreover, adiabatic particle acceleration processes arising in the dipolar magnetic field of the planet (like betatron and fermi acceleration) are unaltered by the planet rescaling as well. This is particularly important for the discussion in Sec.~\ref{sec:Result_electrons} on the loss-cone angle around the planet.

Second, scaling down the planet radius can have an impact on magnetic reconnection in the tail in the case of southward IMF. Magnetic reconnection induces the formation of a region of drifting plasma close to the neutral point, the so-called diffusion region, where particles are no more frozen-in the magnetic field. Depending on the species, this region is called the ion (electron) diffusion region and it extends for about ten ion (electron) inertial lengths around the neutral point.
By scaling down the planet radius, the ion (electron) diffusion region in the tail moves closer to the planet surface. 
In our case, given the scaled down planet we are using ($R=5.5 d_i$, where $d_i\equiv c/\omega_{pi} \equiv c\sqrt{m_i/4\pi n_{_{SW}}e^2}$ is the ion inertial length), the ion (electron) diffusion region is marginally (largely) separated from the planet. Thus we expect ion dynamics in the outflow to be affected by the scaling down, at least partially, while this should not be the case for electrons. In particular, we expect the signatures of high-energy electrons observed in our simulations as a product of magnetic reconnection to hold true when using a real size planet.

Although we expect these two classes of plasma processes to be well-reproduced using the scaled-down parameters in our simulations, particular care must be taken when analyzing specific plasma quantities. Indeed, due to the rescaling of the plasma-to-cyclotron frequency ratio, electrons with energies above few tens of keV fall out of the range of validity of our model since their speed becomes ultra-relativistic (by reducing the ratio $\omega_{pe}/\omega_{ce}$ we are \textit{de facto} reducing the value of the light speed $c$ in our simulations). Moreover, by scaling down the ion-to-electron mass ratio, we reduce the ratio between the electron thermal speed and the flow speed ($v_{the}/V_{_{SW}}$) from 5 to around 1.2. This means that the solar wind electrons interacting with the planet in our simulations are injected with a slightly more anisotropic energy distribution function as compared to the real case. The relevance of these caveats on our simulation results remains unknown and should be further investigated in future works. 
We stress that global fully-kinetic simulations of planetary magnetospheres using realistic plasma parameters remain prohibitive using the present state-of-the-art HPC facilities. 

Since this work represents a first step towards a fully-kinetic global modeling of the Hermean environment, we choose to use a simple as possible realistic solar wind configuration. Thus the choice of a purely northward or southward IMF. This choice facilitates the analysis of the reconnection sites and outflows. A more realistic configuration would require a strong $B_x$ component in the IMF. Such in-plane component, typically found at Mercury, tends to create a foreshock region (absent in our simulations) and drives strong north-south asymmetries in the magnetosphere. Such a more realistic modeling will be addressed in future works.

To conclude, the results of this work show that high energy electrons (up to tens of keV) are generated in the magnetotail of Mercury in the case of southward IMF. Such electrons form as a consequence of magnetic reconnection in the tail neutral line located at $X\approx$-2 $R$ and move towards the planet drifting dawn-ward. When reaching the internal shells of the magnetosphere the motion of these electrons (which can be treated as adiabatic) is strongly affected by loss-cone precipitation on the planet surface. We find that almost all drifting electrons are lost on the planet before completing a full longitudinal drift orbit. Therefore, the highest concentration of electrons in the range of hundreds of eV to some keV is in the nightside, especially in the post-midnight sector.
This result supports electron observations by Mariner10/PLS instrument, showing high energy electrons close and after closest approach inside the magnetosphere. Such \textit{in-situ} observations are in agreement with our simulation with southward IMF. Moreover, we present synthetic electron data useful for the ongoing and future interpretation of BepiColombo/MEA first-Mercury-flyby observations. A detailed comparison between our simulations and MEA observations for this flyby will be done in the near future once the data will be fully calibrated. Finally, we envision a characterization of this energetic electron population by BepiColombo during its nominal science phase.

\begin{acknowledgements}
This work was granted access to the HPC resources at TGCC under the allocation AP010412622 and A0100412428 made by GENCI via the DARI procedure.
We acknowledge the CINECA award under the ISCRA initiative, for the availability of high performance computing resources and support for the project IsC93.
We acknowledge the support of CNES for the BepiColombo mission.
Part of this work was inspired by discussions within International Team 525: ``Modelling Mercury's Dynamic Magnetosphere in Anticipation of BepiColombo'' at the International Space Science Institute, Bern, Switzerland.
We acknowledge support by ESA within the PhD project ``Global modelling of Mercury's outer environment to prepare BepiColombo''.
Analysis of BepiColombo and Mariner10 data was possible thanks to the AMDA (http://amda.irap.omp.eu) and CLWeb (http://clweb.irap.omp.eu) tools.
\end{acknowledgements}


\bibliographystyle{aa} 
\bibliography{biblio}


\end{document}